\documentclass[fleqn,usenatbib]{mnras}

\usepackage{newtxtext,newtxmath}

\usepackage[T1]{fontenc}

\DeclareRobustCommand{\VAN}[3]{#2}
\let\VANthebibliography\thebibliography
\def\thebibliography{\DeclareRobustCommand{\VAN}[3]{##3}\VANthebibliography}

\usepackage[spanish, english]{babel,varioref}
\usepackage{orcidlink}
\usepackage{graphicx}	
\usepackage{amsmath}	
\usepackage{multicol}        
\usepackage{bm}		
\usepackage{pdflscape}	
\usepackage{ulem} 
\usepackage{natbib}
\usepackage{tabularx}
\usepackage{longtable}
\usepackage{float}
\usepackage[flushleft]{threeparttable}
\usepackage{adjustbox}
\usepackage[T1]{fontenc}
\usepackage{ae,aecompl}
\usepackage{xspace}
\usepackage{gensymb} 
\usepackage{subfig}

\newcommand{\hii}{H\,\textsc{ii}}

\newcommand{\oii}{[O\,\textsc{ii}]}
\newcommand{\oiii}{[O\,\textsc{iii}]}

\defcitealias{Rodriguez2020}{R20}
\defcitealias{ChornayWalton2021}{CW21}
\defcitealias{GonzalezSantamaria2021}{GS21}
\defcitealias{DelgadoInglada2015}{DI15}
\defcitealias{MendezAmayo2022}{MD22}
\defcitealias{Spitoni2021}{SP21}

\title[Gradients comparison in the Galactic disk]{A comparative study of O, Ne, Cl, and Ar in \hii~regions and PNe of the Galactic disk: Temporal evolution of radial gradients?}

\author[Amayo et al.]
{A. Amayo,$^{1}$\orcidlink{0000-0002-2915-2425}, 
L. Carigi$^{1}$ \thanks{Contact e-mail: \href{mailto:carigi@astro.unam.mx}{carigi@astro.unam.mx}}\orcidlink{0000-0002-2023-466X}, 
J. E. M\'endez-Delgado$^{1}$\orcidlink{0000-0002-6972-6411}, 
J. García-Rojas$^{2,3}$\orcidlink{0000-0002-6138-1869}, 
E. Reyes-Rodr\'iguez$^{2,3}$\orcidlink{0000-0003-1192-6987}\\
$^{1}$Universidad Nacional Autónoma de M\'exico. Instituto de Astronom\'ia. A.P. 70-264, 04510. Ciudad de M\'exico, M\'exico.\\
$^{2}$Instituto de Astrofísica de Canarias, E-38205 La Laguna, Tenerife, Spain. \\ $^{3}$Departamento Astrofísica, Universidad de La Laguna, E-38206 La Laguna, Tenerife, Spain.\\
}
\date{Last updated ...; in original form ...}
\pubyear{2025}

\begin{document}
\label{firstpage}
\pagerange{\pageref{firstpage}--\pageref{lastpage}}
\maketitle

\begin{abstract}
We compare the radial abundance gradients of O, Ne, Cl, and Ar using a sample of 42 \hii\ regions and 176 planetary nebulae (PNe) from the DESIRED catalogue in the Galactic disk. Both samples comprise the highest-quality observations currently available. Actually, this work presents the first gradient analysis for the DESIRED dataset. For all objects, two sets of chemical abundances were compiled: one derived from collisionally excited lines (CELs) and another incorporating the temperature fluctuation parameter ($t^2$). Oxygen abundances were corrected for dust depletion, and the results for \hii\ regions were compared with those of Cepheid stars, which trace the present-day interstellar medium. Distances for PNe were carefully compiled from the most recent literature, excluding objects associated with the bulge or halo to ensure a disk-only sample. All gradients are statistically significant ($p$-values $<0.05$), except for Cl and Ar in \hii\ regions. The O/H gradients derived from \hii\ regions and Cepheids are consistent when $t^2$ is included, underscoring the importance of accounting for temperature inhomogeneities in nebular analyses. 
The O and Ne gradients traced by older objects are flatter than the present-day gradient by $-0.028\pm0.008$ dex kpc$^{-1}$ on average, from both RLs and CELs. This could be interpreted as a temporal steepening of the Galactic abundance gradient in ISM; however, such behavior is not reproduced by several chemical evolution models, suggesting that additional physical processes could be influencing the observed trends. The most plausible explanation is that our PNe sample has been strongly affected by radial migration. Under this interpretation, the PNe gradient cannot reliably trace past abundance gradients, but it provides a valuable constraint on radial stellar migration, offering important input for chemo-dynamical models of the Galactic disk and for hydrodynamical simulations.

\end{abstract}




\section{Introduction}


The Milky Way provides a unique opportunity to conduct detailed chemical studies of both stellar and nebular objects, serving as an exceptional laboratory for refining and constraining chemical evolution models. Understanding the chemical history of our own Galaxy represents a benchmark for studies of the chemical evolution of other galaxies and constitutes a fundamental step towards understanding the evolution of the Universe \citep[e.g.,][]{Chiappini1997, Pilkington2012, Kubryk2015, Carigi2019, Palla2020, Vincenzo2020, SpitoniRecioBlanco2023}.

The Galactic chemical radial gradients, traced through either nebular or stellar objects, encompass populations spanning a wide range of ages. The patterns of these distributions, as in other galaxies, reflect the star formation history, the galactic mass, and the efficiency of gas inflows and outflows events \citep[e.g.,][]{Tinsley1980, Pilkington2012, Molla2015}. Depending on the age of the studied objects, the radial gradients reveal the chemical conditions at different epochs of the Galactic history. For instance, \hii~regions probe the present-day chemical composition of the interstellar medium (ISM), while planetary nebulae (PNe) provide information on past chemical conditions. 

Among all metals, oxygen is a benchmark element for studying the chemical evolution of the Milky Way based on \hii~regions and PNe. In these nebulae, \oii~and \oiii~emission lines --corresponding to the O dominant ionic species, O$^{+}$ and O$^{++}$-- are very bright and easily detectable with modern instruments. Galactic O distributions in nebulae can be determined through different methods, including optical spectroscopy \citep{Esteban2018, arellano2020}, infrared observations \citep{Rudolph2006}, and radio measurements \citep{Shaver1983, Wenger:2019}. On the other hand, stellar oxygen determinations are considerably more challenging for several reasons: the intrinsic weakness of oxygen features, the complexity of stellar atmospheres, the presence of multiple ionization stages, and the need for accurate stellar models \citep[see][ for a comprehensible review]{Jofre2019, SimonDiaz2020}. 

Previous studies of the Galactic oxygen gradient with \hii~regions found a consistent negative slope across a wide range of Galactocentric distances. The slope of the O gradient typically ranges from $-0.040$ to $-0.060$ dex kpc$^{-1}$  \citep{Shaver1983,Deharveng2000,Rudolph2006, Balser2011, arellano2021}, with significant dispersion in some datasets \citep[see discussion in][]{MendezAmayo2022}. This result supports the \textit{inside-out} formation scenario \citep[e.g.,][]{Chiappini1997, Hayden2015, Frankel2019}, in which the star formation rate (SFR) was more efficient at early times in the inner regions of the Galaxy, $3\leq R \leq 5 $ kpc (excluding the bulge region), than in the outer regions.

The study of Galactic gradients based on PNe is a more challenging task, as it introduces an explicit time dependence. Depending on the PNe ages, the slopes of the O gradient can range from $-0.072$ to $-0.014$ dex kpc$^{-1}$ \citep[see][and references therein]{Perinotto2006}. When the ages of PNe are considered in detail, these objects allow the investigation of the temporal evolution of the O gradient. However, previous studies have led to contradictory results. For instance, \citet{Maciel2003} found that younger PNe (with ages from 0 to 3 Gyr) exhibit a flatter O gradient than older ones (with ages greater than 6 Gyr), by up to 0.05 dex kpc$^{-1}$, while \citet{Stanghellini2010} reported an increase in the slope of the O gradient of 0.024 dex kpc$^{-1}$ from the past to the present \citep[see][for a brief summary on this matter]{GarciaRojas2019}. Extragalactic studies in other disk galaxies (e.g., M~31, M~33, M~81, NGC~300), lead in most cases to a steeper gradient in younger objects than in older ones, however, they also find a much more considerable dispersion in old objects, suggesting the latter are less reliable and that other physical processes may alter the original gradients in time \citep[e.g.,][]{Magrini2007, Magrini2016, Stasinska2013, Stanghellini2014, Arnaboldi2022}. 

In stellar studies, iron is commonly used as a proxy for metallicity due to its spectral accessibility. Many studies investigating the Galactic chemical gradient from stars rely on Fe abundances. Using APOGEE DR16 data, \citet{Lian2023} identified a change in slope in the light-weighted integrated stellar metallicity profile at $R=6.9 \pm 0.06$ kpc. Inside this radius, the slope is positive ($0.031 \pm 0.010$ dex kpc$^{-1}$), while beyond it, the slope becomes negative ($-0.052 \pm 0.008$ dex kpc$^{-1}$). However, this breakpoint disappears when stars are categorized by age ranges, with steep and negative gradients observed in younger stellar populations and flat gradients in the older ones. The steep gradient found in younger populations aligns with results from young stars and \hii~regions in the Milky Way. Nonetheless, the comparison between stellar Fe gradients and nebular O gradients is not straightforward, as it depends on the variation of O/Fe with time and environment \citep{Asplund:05, Amarsi:19, Chruslinska:24, MendezDelgado2024b}. Furthermore, iron is heavily depleted into dust in photoionized environments, complicating the estimation of its total abundance \citep{Rodriguez:02, Izotov:06, MesaDelgado2009, MendezDelgado:21a, MendezDelgado2024b}. Concerning temporal evolution of gradients, some stellar studies suggest no evolution or a mild change in slope from the past to the present \citep[e.g.,][]{Magrini2009, Willett2023}, but is still an open question in literature.

Elements such as Ne or Ar belong to the group of $\alpha$-elements, together with O. These elements are synthesized through $\alpha$-capture nucleosynthesis in massive stars, and their abundance ratios are expected to remain nearly constant \citep{Vincenzo&Kobayashi2018, Kobayashi:20, Esteban2025}. Although chlorine is classified as a halogen, its production is linked to that of Ar \citep{Clayton:03}, suggesting a behaviour similar to that of the $\alpha$-elements. However, it has been proposed that some planetary nebulae (PNe) may produce O during their lifetime \citep[][]{DelgadoInglada2015}, unlike Ar, Ne, and Cl. Emission lines from these elements are present in the optical spectra of photoionized nebulae, and studying their abundances may play a crucial role in understanding the Galactic chemical evolution, particularly if some O is indeed produced in PNe. Furthermore, Ar and Ne are noble gases and are not expected to form dust compounds that could trap O. Moreover, as noble gases, Ar and Ne are not expected to be incorporated into dust, unlike oxygen, which can be depleted into solid compounds \citep{Peimbert2010}.

Regarding the determination of heavy element abundances in ionized nebulae, one major aspect to consider is the long-standing abundance discrepancy problem (AD). This problem, first reported by \citet{Bowen:1939} and \citet{Wyse:1942}, has been extensively studied \citep[e.g.,][]{Peimbert1967,Torres-Peimbert:1990,Liu:2000,garcia-rojas:2007b,Nicholls:2012,MendezDelgado2023}. It consists of a systematic difference between the abundances derived from heavy element recombination lines (RLs) and those obtained from collisionally excited lines (CELs), with RL-based abundances consistently higher than CEL-based ones. This difference is commonly quantified by the abundance discrepancy factor (ADF) \citep[e.g.,][]{tsamisetal03}.

If temperature inhomogeneities are assumed to be the main cause of the ADF, as proposed by \citet{Torres-Peimbert:1980}, the so-called root mean square temperature fluctuation parameter \citep[$t^2$][]{Peimbert1967} can be incorporated into the calculation of chemical abundances. This parameter quantifies the temperature inhomogeneities and corrects their systematic effects. Observational evidence supports the presence of temperature inhomogeneities within \hii~regions \citep{Peimbert:2003,garcia-rojas:2007b, Peimbert:2013, MendezDelgado2023}. However, the AD problem in planetary nebulae (PNe) appears to be more complex. In many of these objects, temperature variations alone cannot explain the observed ADF, which can reach values higher than a factor of 50 in some cases\footnote{For a compilation of the ADF measured in photoionized nebulae visit: \href{https://nebulousresearch.org/adfs/}{https://nebulousresearch.org/adfs/}.} \citep{Wesson2018, GarciaRojas2022}. For PNe with extreme ADFs, the inclusion of high-metallicity clumps has been proposed as an alternative explanation \citep{Torres-Peimbert:1990, Liu:2000, GarciaRojas2022, Richer:2022}. However, these clumps introduce additional complications, as the volumes emitting RLs and CELs differ significantly with little interrelation \citep[see discussion in][]{MendezDelgadoproceeding:2023}.

Within the framework of chemical evolution models, \citet{Carigi2019} explored the impact of the AD by computing two models. The first model was designed to reproduce the O/H abundances of \hii~regions derived from RLs, while the second model aimed to reproduce those obtained from CELs. They found that the first model successfully reproduced stellar constraints based on B-stars, Cepheids, and the Sun, whereas the second model failed to do so. Furthermore, \citet{MendezAmayo2022} investigated how the inclusion of $t^2 > 0$ in abundances originally derived from CELs affects the chemical gradients of \hii~regions. They observed increases of up to 0.3 dex in the gradients compared to those derived with $t^2 = 0$ (in absolute abundance values, not the slopes). This difference led to more consistency in the gradient slopes of $\alpha$-elements (O, Ne, and Ar).

In this paper, our main aim is to study the Galactic radial chemical gradients of O, Ne, Cl and Ar, and their temporal evolution, in the light of recent and reliable available data of stars, PNe and \hii~regions. We also investigate how the Galactic chemical gradients of PNe and \hii~regions differ when abundances are derived using CELs or CELs+$t^2$ (RL-like) methods. The observational sample, consisting of \hii~regions, Cepheids (used for comparison with the \hii~regions) and PNe (previously classified in age intervals) is presented in Section~\ref{sec:observations}. The resulting chemical radial gradients are presented in Section~\ref{sec:ChemGrads}, and are discussed in Section~\ref{sec:Discusion}. Finally, our conclusions are shown in Section~\ref{sec:Conclu}

\section{Observational sample}
\label{sec:observations}
\subsection{\hii~regions}
\label{subsec:r_hiis}

We adopt the chemical abundances and Galactocentric radial distances for a sample of 42 Galactic \hii~regions, compiled from the works of \citet{arellano2020,arellano2021}, and \citet{MendezDelgado2020}, as presented in \citet[][hereafter MD22]{MendezAmayo2022}, which constitute a carefully curated dataset of reliable objects in the Milky Way. These studies are based on strictly homogeneous analyses and have been progressively expanded with high-quality observations, yielding the lowest-dispersion Milky Way abundance gradients reported to date. The objects in this sample possess the highest quality optical spectra currently available, obtained with 8-10m class telescopes at intermediate-to-high spectral resolution (up to R = $\lambda/\Delta\lambda \sim 10,000$). Compared to other literature compilations, this sample shows significantly reduced scatter at a given Galactocentric distance \citep[see e.g., Section 5.1 of ][]{arellano2020}, reinforcing their suitability for precise gradient determinations.

Besides the careful and homogeneous determination of the chemical abundances of He, C, N, O, Ne, S, Cl, and Ar, \citetalias{MendezAmayo2022} also present Galactocentric distances, primarily using Gaia EDR3 parallaxes of the ionizing stars. After comparing distances determined through various methods, these authors adopted a final set of distances, which we use in this work. Additionally, \citetalias{MendezAmayo2022} applied a mean global temperature fluctuation parameter ($t^2$, \citealt{Peimbert1967}) greater than zero to the ionic abundances of N, O, Ne, S, Cl, and Ar in order to examine its impact on the Galactic radial gradients. They found that assuming $t^2=0.038\pm0.008$ seems to resolve the discrepancies previously reported by \citet{arellano2020,arellano2021} between the nebular gradients and the solar abundances (either photospheric or derived from the solar wind) reported by \citet{Lodders2019} and \citet{Asplund2021}.

In this paper, we gather both sets of O, Ne, Cl, and Ar abundances from \citetalias{MendezAmayo2022}, namely those derived under the assumption of $t^2 = 0$ and those corrected with $t^2 > 0$. In addition, both O abundances have been corrected for dust depletion by $\sim$0.1 dex,
as described in Section~\ref{subsec:dustcorr}. 

\subsection{Cepheids}
\label{subsec:Cepheids}


It is possible to compare the O gradients derived from \hii~regions with those obtained from classical Cepheids. These stars are luminous objects, observable at large distances due to their high brightness. Their spectra contain numerous absorption lines from different elements, enabling accurate determinations of their chemical composition \citep[e.g.,][]{Lemasle2013}. Classical Cepheids are very young \citep[with ages below 200 Myr, see][]{Bono2020} and their O abundance is not expected to be nucleosynthetically modified \citep{Luck2008}. This makes them directly comparable to \hii~regions and thus, their abundance results are expected to be consistent. 

We adopted the oxygen abundances from the sample of 435 Cepheids studied by \citet{Luck2018}. This work updated earlier determinations of Cepheid chemical abundances from high-resolution observations presented in \citet{Andrievsky2002a, Andrievsky2002b, Andrievsky2002c, Andrievsky2004, Andrievsky2005, Luck2003, Luck&Andrievsky2004, Kovtyukh2005, Luck2006, Luck2008, Luck&Lambert2011, Luck2011, Luck2014}, and applying a homogeneous methodology. These updated compilation nearly doubled the previous stellar sample from \citet{Luck&Lambert2011} and incorporated the most up-to-date physical considerations.


The Cepheids sample from \citet{Luck2018} includes both classical Cepheids and some type II Cepheids, the latter potentially reaching ages up to 10 Gyr \citep{Bono2020}. To focus exclusively on young Cepheids with chemical abundances comparable to those of \hii~regions, we restricted our analysis only to the classical Cepheids. For these stars we adopted the oxygen abundances reported by \citet{Luck2018}, assuming an average uncertainty of 0.1 dex, based on their quoted errors in [O/Fe]. We gathered the geometric distances of the Cepheids sample, obtained by \citet{BailerJones2021}, using Gaia DR3 parallaxes. Following the criteria established by \citet{Luri2018} and \citet{Stanghellini2020}, we retained only stars with parallax uncertainties below 20 per cent. Although this selection criterion could potentially bias the distance range spanned by the Cepheid sample, it is unlikely to do so, as the final sample is distributed between 4.2 and 14.7 kpc from the Galactic centre, a range wide enough to allow a robust study of the Galactic disk. With this Cepheid sample, we computed the O/H gradient and compared it with that of the \hii~regions (which are located in the range $5 < R_{G} < 17$ kpc) representing the present-day ISM.

\subsection{Planetary nebulae sample}
\label{subsec:pns}

The PNe sample is drawn from the DEep Spectra of Ionized Regions Database  \citep[DESIRED;][]{MendezDelgado2023}. This database compiles high signal-to-noise optical spectra of photoionized nebulae, providing detailed emission-line measurements across a wide wavelength range and enabling accurate determinations of physical conditions and chemical abundances. Some of the most recent studies based on this sample can be found in \citet{MendezDelgado2024, Esteban2025, OrteGarcia2025} and \citet{MendezDelgado2025}.
PNe in the DESIRED catalogue represents one of the most extensive and homogeneous compilations of planetary nebula data currently available. It integrates spectroscopic observations from a wide range of literature sources, yet rederives chemical abundances through a uniform methodology, ensuring internal consistency across the entire dataset. This homogeneity is a key strength, as it minimizes systematic discrepancies arising from different analysis techniques, allowing for robust comparative studies of elemental gradients and abundance patterns. The large sample size further enhances its statistical power, enabling detailed investigations of the chemical evolution of the Milky Way with unprecedented robustness. 
The DESIRED methodology is drafted in \citet{MendezDelgado2023b}. The only difference with PNe is the use of the ionization correction factor (ICF) scheme by \citet{DelgadoInglada2014}. The assumed temperature structure is a three-zone scheme. Despite its completeness and quality, no dedicated studies of distances—or, even more so, of chemical abundance gradients—have been carried out for these PNe until the present work.

At the time of access, the DESIRED database contained 3,154 spectra from various bibliographic sources, covering the Milky Way and different galaxies from the local group to high-redshift systems, and focused on {\hii} regions and star-forming regions. By selecting PNe located in our Galaxy, the sample was reduced to 229 objects. For this initial selection, we excluded PNe from \citet{Tan2024}, since these objects belong to the Galactic Bulge.

Among the selected objects, twelve PNe (Hf~2-2, M~1-11, NGC~1501, NGC~2440, NGC~2867, NGC~3242, NGC~6439, NGC~650, NGC~6620, NGC~6778, NGC~6884, and PB~6) have two different sets of spectra/abundances, corresponding to either different bright regions within the same object (NGC~2867, NGC~3242, NGC~650, and PB~6) or to spectra published in different references (the remaining cases). A detailed study of the dispersion introduced by different datasets for the same object is beyond the scope of this paper \citep[for a detailed analysis of this dispersion, the reader is referred to][]{Rodriguez2020}. We therefore adopted a single spectrum for each of the PNe with more than one abundance set as follows. 
For Hf~2-2 we adopted the abundances resulting from the spectra of \citet{Wesson2018}, since the alternative spectra \citep{GarciaRojas2022} do not cover the important auroral \oiii\ $\lambda$4363 line. For NGC~1501, NGC~6439, NGC~6620, and NGC~6884, we select the abundances from the spectra of \citet{Ercolano2004, WangLiu2007}; \citet{WangLiu2007} and \citet{Liu2004} respectively, based on their reported uncertainties in O/H. In any case, the O/H values from different references agree within uncertainties. For M~1-11, NGC~2440, and NGC~6778, only one spectrum provided Ar abundances (an element of interest in the present study), and in such cases, we selected that spectrum \citep[those from][respectively]{Henry2010, Sharpee2007, GarciaRojas2022}. Finally, for the remaining PNe with more than one reference (NGC~2867, NGC~3242, NGC~650, and PB~6), we compute mean abundances of O, Ne, Cl, and Ar across the available datasets.
Differences in O/H between datasets for this group are small and within uncertainties ($0.02$--$0.10$ dex), which suggests that this approach is acceptable and does not introduce significant systematic effects. 

For \hii\ regions, the general methodology for computing physical conditions and chemical abundances —including the temperature and density structure and the ICF scheme— followed the procedures described in \citet{MendezDelgado2023b}. In the case of PNe, we adopted the same temperature/density framework, while elemental abundances were calculated using the ICF scheme proposed by \citet{DelgadoInglada2014}.

\subsubsection{Galactocentric distances}
\label{subsubsec:PNeDists}

Once the abundances for each object had been determined and a single spectrum was selected for each nebula, we gathered the distances reported in the detailed analysis of \citet{HernandezJuarez2024}.\\

\citet{HernandezJuarez2024} present a new catalogue of distances for 2,211 PNe, derived from both Gaia parallaxes and statistical estimations. The authors developed a method to identify the most reliable distance for each PN, accounting for known Gaia limitations such as systematic errors and spurious solutions. Distances were computed using three statistical catalogues (\citealt{Zhang1995, Frew2016} and \citealt{Stanguellini2018}) in combination with Gaia EDR3 data, considering only positive parallaxes with a good astrometric solution (represented by a Renormalised Unit Weight Error, RUWE, lower than 1.4). Priority was given to distances derived from the inverse of the parallax when the relative error was below 15 per cent, otherwise adopting the median of the available estimates. This homogeneous approach yields improved reliability compared to heterogeneous literature values, with 43 per cent of the distances coming directly from Gaia and the remainder from statistical determinations, all with uniformly assigned uncertainties. Although the method remains subject to the intrinsic limitations of parallax measurements --particularly for distant or compact PNe-- it provides one of the most robust and internally consistent sets of distances currently available. These characteristics make the \citet{HernandezJuarez2024} catalogue a well-justified choice for the present work, ensuring both accuracy and methodological consistency in the adopted Galactocentric distances.

From the cross-match of the \citet{HernandezJuarez2024} catalogue and our DESIRED-Galactic sample of PNe, we obtain Galactocentric distances for all objects, except 11 PNe (BV5-2, H~1-35, He~2-15, IC~1257, IC~5217, M~1-11, M~1-14, M~1-17, M~1-57, MPA~1759, and NGC~6803), which were not studied by \citet{HernandezJuarez2024}. For these objects, we searched for Gaia EDR3 data, using the Gaia IDs reported consistently by \citet{ChornayWalton2021} and \citet{GonzalezSantamaria2021}. We then obtained their Galactocentric distances from the catalogue presented by \citet{BailerJones2021}. We also recompiled their statistical distances from the scale-relation by \citet{Frew2016}. Among these objects, only M~1-11 have a reliable Gaia EDR3 astrometric solution, as indicated by its parallax error lower than 15 per cent. Therefore, for this PN we adopted the distance from \citet{BailerJones2021}, while for the rest we relied on the distances from \citet{Frew2016}.
For He~2-15, IC~1257, M~1-17, and MPA~1759 no Gaia ID was reported by either \citet{ChornayWalton2021} or \citet{GonzalezSantamaria2021}, and no distance was available in \citet{Frew2016}. These objects were thus discarded from the present sample, reducing our set to 216 PNe.   

For this sample, the Galactocentric distances ranged from 0.4 kpc to 26 kpc. In the literature, most criteria to identify whether a PN belongs to the bulge are based on Galactic coordinates (longitude $0$\degree $\leq l \leq 10$\degree or $350$\degree $\leq l \leq 360$\degree\ and latitude $-10$\degree $< b < 10$\degree), angular diameter (less than 20 arcseconds), and flux at 5 GHz (less than 100 millijansky) \citep[e.g.,][]{Bensby2001, Stanghellini2012}. Following the approach of \citealt{Brayan2020}, we assumed a spherical bulge with an angular size of 10\degree\, located at $8.20\pm0.10$ kpc \citep{BlandHawthorn2016}, leading to a minimum Galactic radius of $1.45\pm0.02$ kpc for an object to belong to this Galactic component (a value very similar to those obtained by \citealt{Bensby2001}).
Based on this criterion, we identified five PNe that are highly likely to belong to the Galactic bulge due to their Galactocentric distances being smaller than $1.45\pm0.02$ kpc: H~1-40, H~1-54, Hen~2-283, M~2-33, and M~2-39, with distances of $0.432\pm{1.416}$, $0.920\pm{2.388}$, $1.047\pm{2.526}$, $0.936\pm{0.188}$, and $1.436\pm{1.209}$ respectively. Consequently, these objects were removed from the sample, resulting in a total of 211 disk PNe.

The working sample built under the above criteria spans Galactocentric distances from 1.46 kpc to 25.78 kpc. Most observational evidence indicates that the Galactic halo extends up to radii beyond 100 kpc, with a significant density break at approximately $R_{G}\sim15-30$ kpc, where halo stars become dominant \citep{Xue2015, Amarante2024}. While some disk stars have been detected at larger radii, such cases are expected to be rare \citep[][]{LopezCorreidora2018}. The Galactic thick disk is substantially more radially concentrated, showing a steep decline beyond $R_{G}\lesssim10$ kpc \citet{Bensby2017, Katz2021}. However, the thin disk extends farther outward, with an observed radial scale length of up to $\sim12$ kpc, beyond which the stellar density declines significantly \citet{LopezCorreidora2018, Katz2021}. These studies suggest that adopting a conservative cutoff at $R_G=12$ kpc could minimize contamination from halo objects. However, our aim is to work with the most extensive and consistent sample possible within the radial range probed both by \hii~regions and PNe. Based on this, we adopt a radial cutoff at $R_{G}\leq17$ kpc for our abundance-gradient analysis, matching the maximum radius reached by the \hii~region sample. We further applied a vertical cut of $|Z| \leq 1.0$ kpc, which excludes regions where the stellar halo becomes significant while retaining both thin- and thick-disk objects \citep[e.g.,][]{Zijlstra1991, Jong2010, Bucciarelli&Stanghellini2023}. 

These radial and vertical selections yield a final sample of 176 PNe, located within the Galactic disk from $R_G=1.46$ kpc to $\sim17$ kpc, with a mean height relative to the plane of $|Z| = 0.36$ kpc.

\subsubsection{Oxygen depletion correction}
\label{subsec:dustcorr}

When dealing with gaseous oxygen abundances, both in \hii~regions and PNe, it is crucial to consider that a portion of the atoms may be incorporated into dust grains, which can survive in photoionized environments. However, the mechanisms governing dust formation \citep[e.g.,][]{Gehrz1989}, destruction \citep[e.g.,][]{Rodriguez:02, Rodriguez:05}, and evolution \citep[e.g.,][]{Kwok1980, Stanghellini2012} in these environments remain open questions. The physical conditions of the formation environment, together with the evolution of the dust, affect its composition \citep{Jones&Ysard2019}, leading to expected differences among the dust present in \hii~regions and in PNe. 

In the case of \hii~regions, we correct oxygen abundances using the upper limits to the depletion factors computed by \citet{Peimbert2010}: $0.09\pm0.01$ dex for objects with $7.3 < 12+\log$(O/H) $< 7.8$; $0.10\pm0.01$ dex when $7.8 < 12+\log$(O/H) $< 8.3$; and $0.11\pm0.01$ dex for $8.3 < 12+\log$(O/H) $< 8.8$. However, caution is required when using these values, since they were obtained based on: i) the oxygen depletion computed by \citet{MesaDelgado2009} for the Orion Nebula, ii) the cosmic standard abundances derived by \citet{Przybilla2008} from B-type stars, iii) the solar oxygen abundance of \citet{Asplund2009}, and iv) the oxygen, magnesium, silicon, and iron abundances measured in the Orion Nebula, together with the adopted $t^2$ parameter. These reference values may now be somewhat outdated and merit re-evaluation. Additionally, it is important to highlight that the oxygen-depletion corrections proposed by \citet{Peimbert2010} represent upper limits, since the fraction of oxygen trapped in dust is expected to decrease as the ionizing field becomes harder, an expected situation for low-metallicity environments, as extensively discussed by \citet{MendezDelgado2024b}.

For PNe, the dust correction is even more uncertain. Several studies in the literature assume that oxygen is predominantly incorporated into silicates and oxides within the dust of these objects \citep[e.g.,][]{Meyer1989, Cardelli1996, MesaDelgado2009}. Based on this assumption, we adopt a uniform correction of 0.10 dex to the O abundances in our PN sample.

\section{Radial gradients of O, N\lowercase{e}, C\lowercase{l}, and A\lowercase{r}}
\label{sec:ChemGrads}

Using the chemical abundances and Galactocentric distances of our \hii~region and PN samples, we calculated the radial gradients of O, Ne, Cl, and/or Ar, obtained from CELs and RLs (or CELs corrected by t$^2$). In the case of \hii~regions, we include both RLs and CELs$+t^2$; however, for simplicity, we will collectively refer to them as the RL abundances, despite their methodological differences.

For each sample, the radial gradient was computed using the linear relation $y = a + b R_G$, where $a$ and $b$ correspond to the intercept and the slope, respectively. This functional form assumes no significant change in the slope at any particular radius --a possibility discussed in earlier works \citep[e.g.,][]{Esteban2018, Vilchez1996, Maciel2010}, but not supported by more recent observations \citep{arellano2020, arellano2021}.

The coefficients $a$ and $b$ were obtained through a \textit{Monte~Carlo} procedure that accounts for the uncertainties in both $x$ and $y$, as well as for the intrinsic scatter of the data. 
The resulting median values of $a$ and $b$ and their $1\sigma$ uncertainties are presented in Table~\ref{tab:GradObs}.

As mentioned above, each sample spans a slightly different $R_G$ interval: \hii~regions (42 objects) cover $5 \leq R_G \leq 17$ kpc, while PNe (176 objects) extend over $1.46 \leq R_G \leq 16.43$ kpc, making the two datasets statistically comparable. We remark that these samples trace different epochs in the chemical evolution of the Galactic disk: \hii~regions represent the present-day ISM whereas PNe trace the ISM composition from approximately 2--4 Gyr ago \citep{Maciel2010}.

\begin{figure*}
\centering
\includegraphics[width=\textwidth]{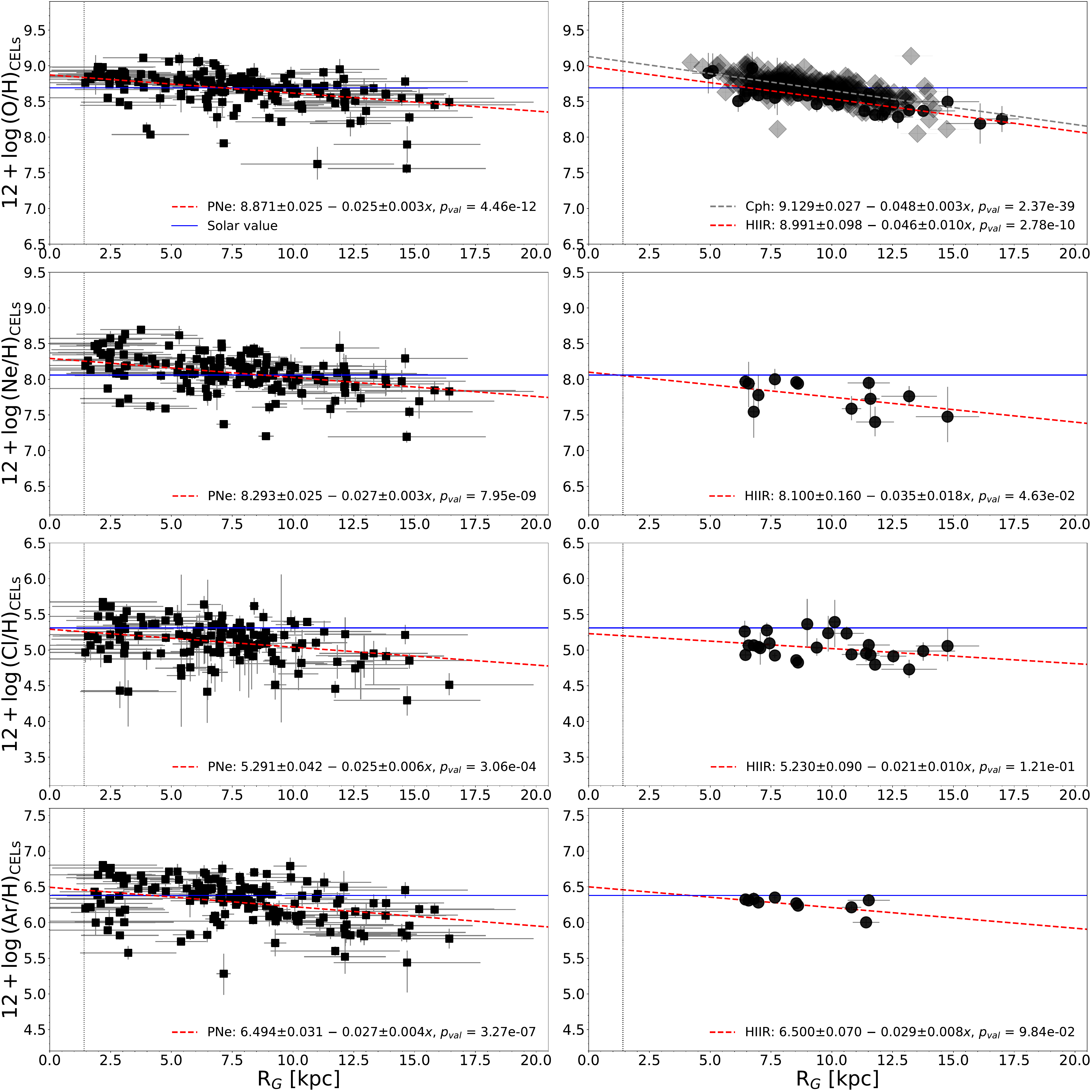}
\caption{Radial abundance gradients of O, Ne, Cl, and Ar computed from CELs for PNe (representing past conditions; left panels) and for \hii~regions and Cepheids (representing present-day conditions, right panels). PNe are shown as black squares, \hii~regions as black circles, and Cepheids as grey diamonds. Linear fits are shown as red dashed lines for PNe and \hii~regions, and as grey dashed line for Cepheids. The solid blue line indicates the solar abundances by \citet{Asplund2021}. The vertical dotted line marks the $R_{G}$ limit considered to exclude bulge objects.}
\label{fig:NewGradCELS}
\end{figure*}

\begin{figure*}
\centering
\includegraphics[width=0.9\textwidth]{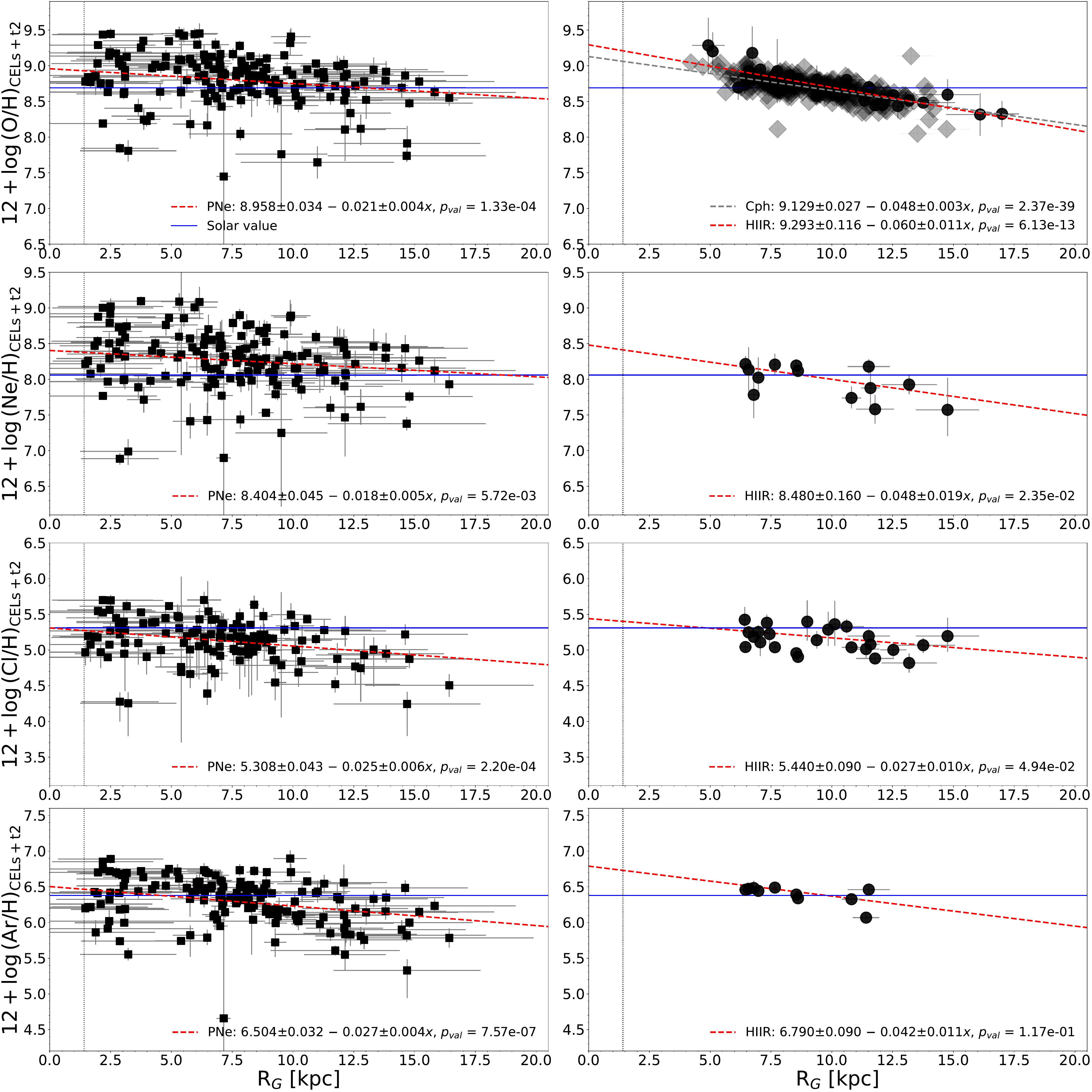}
\caption{Radial abundance gradients of O, Ne, Cl, and Ar computed from RLs (CELs$+t^{2}$) with the same colours and lines as Fig.~\ref{fig:NewGradCELS}.}
\label{fig:NewGradCELS+t2}
\end{figure*}

Figures~\ref{fig:NewGradCELS} and \ref{fig:NewGradCELS+t2} display the resulting gradients. 
Figure~\ref{fig:NewGradCELS} shows the O, Ne, Cl, and Ar gradients based on CELs, whereas Figure~\ref{fig:NewGradCELS+t2} presents those derived from RLs (CELs+$t^2$). In both figures, the right-hand panels illustrate the present-day abundances, with \hii~regions shown as black circles and Cepheids as grey diamonds, together with their regression lines in red and gray dashed lines, respectively. The dotted vertical line marks our adopted cutoff in $R_G$ to exclude bulge objects. The horizontal solid blue line represents the solar abundances from \citet{Asplund2021}. The left-hand panels in both figures illustrate results for the past with PNe (black squares), using analogous regression lines and reference markers. 

We further discuss the implications of these gradients --including the caveats associated with comparing oxygen abundances in PNe and \hii~regions-- in the next section.

\begin{table*}
\centering
\begin{adjustbox}{width=\textwidth}
\begin{tabular}{|lc|cccc|lcccc|}
\hline
                                 & Time range & N   & $a$             & $b$                & $p$-value  &                  & N   & $a$             & $b$              & $p$-value  \\ \hline
O/H$_{\rm CELs}$                 &               &     &                 &                    &            & O/H$_{\rm RLs}$  &     &                 &                  &            \\
Cepheids                        & present       & 325 & 9.13 $\pm$ 0.03 & $-$0.050 $\pm$ 0.003 & $<<$ 0.005 &                  & 325 & 9.13 $\pm$ 0.03 & $-$0.050 $\pm$ 0.003 & $<<$ 0.005 \\
\hii~Regions & present       & 42  & $9.01\pm0.06$   & $-0.05\pm0.01$     & $<< 0.005$ &                  & 42  & $9.32\pm0.08$   & $-0.06\pm0.01$   & $<< 0.005$ \\
PNe                              & past        & 176  & $8.87\pm0.02$   & $-0.024\pm0.003$     & $<<$ 0.005    &                  & 176  & $8.96\pm0.03$   & $-0.021\pm0.004$   & $<<$ 0.005      \\ \hline
Ne/H$_{\rm CELs}$                &               &     &                 &                    &            & Ne/H$_{\rm RLs}$ &     &                 &                  &            \\
\hii~Regions & present       & 13  & $8.13\pm0.18$   & $-0.04\pm0.02$     & 0.046      &                  & 13  & $8.44\pm0.168$  & $-0.05\pm0.02$   & 0.024      \\
PNe                              & past          & 160  & $8.29\pm0.03$   & $-0.027\pm0.004$     & $<<$ 0.005      &                  & 160 & $8.40\pm0.05$   & $-0.018\pm0.005$   & 0.005      \\ \hline
Cl/H$_{\rm CELs}$                &               &     &                 &                    &            & Cl/H$_{\rm RLs}$ &     &                 &                  &            \\
\hii~Regions & present       & 25  & $5.23\pm0.11$   & $-0.02\pm0.01$     & 0.121      &                  & 25  & $5.40\pm0.13$   & $-0.03\pm0.01$   & 0.05       \\
PNe                              & past          & 133  & $5.29\pm0.04$   & $-0.025\pm0.006$     & 0.0003      &                  & 133  & $5.31\pm0.05$              & $-0.025\pm0.006$                & 0.0002         \\ \hline
Ar/H$_{\rm CELs}$                &               &     &                 &                    &            & Ar/H$_{\rm RLs}$ &     &                 &                  &            \\
\hii~Regions & present       & 10  & $6.54\pm0.30$   & $-0.03\pm0.04$     & 0.098      &                  & 10  & $6.76\pm0.53$   & $-0.04\pm0.07$   & 0.117      \\
PNe                              & past          & 158  & $6.50\pm0.03$   & $-0.027\pm0.004$     & $<<$ 0.005       &                  & 158  & $6.50\pm0.03$              & $-0.027\pm0.004$               & $<<$ 0.005          \\ \hline

\end{tabular}
\end{adjustbox}
\caption{Intercepts ($a$), slopes ($b$), and $p$-values of the O, Ne, Cl, and Ar radial abundance gradients, computed from CELs (left column) and RLs or CELs$+t^{2}$ (right column). Cepheids values (available only for O) are shown in boldface in both columns. The "N" columns indicate the number of objects used to compute the corresponding gradient.}

\label{tab:GradObs}
\end{table*}

\section{Discussion}
\label{sec:Discusion}

\subsection{Statistical significance and comparison with literature}

In Table~\ref{tab:GradObs}, most statistical tests yield $p-$values consistently below the conventional threshold of $0.05$. This indicates that the null hypothesis can be rejected with high confidence, implying that most of the observed correlations are unlikely to result from random fluctuations and should therefore be regarded as statistically significant. The only exceptions are the gradients of Cl and Ar in \hii~regions, for which the results should be taken with caution.

The abundance gradients derived for \hii~regions are consistent, within uncertainties, with those reported by \citet{MendezDelgado2022}, despite methodological differences --notably, that study did not apply oxygen dust corrections. The slopes of the O abundance gradients derived from both CELs and RLs of our \hii~regions are consistent, within uncertainties, with those obtained from Cepheids (differing by at most 0.01 dex kpc$^{-1}$). This consistency is expected, as both tracers represent young populations that trace the present-day composition of the ISM. The absolute O abundances are also very similar. However, it can be noticed that O-RLs abundances reproduce those of Cepheids better than those obtained from CELs, supporting the view that adopting the $t^2$ correction (or using RLs directly) provides a more reliable O gradient.

The DESIRED PNe gradients are presented here for the first time, as distance-based analyzes of this sample had not previously been conducted.
They display a coherent pattern across all elements, characterized by moderately negative slopes --a result broadly consistent with previous PNe-focused studies. 
Our O/H gradients derived from CELs and RLs are indistinguishable within the uncertainties. Both are in very good agreement with the gradients obtained for young (Type II) PNe by \citet{Perinotto2006, Stanghellini2010, Maciel2015}, and \citet{Bucciarelli&Stanghellini2023}, whose reported slopes are $-0.02\pm0.01$, $-0.023\pm0.005$, $-0.025\pm0.006$, and $-0.022\pm0.008$ dex kpc$^{-1}$, respectively. This agreement is expected, given that our PNe are estimated to be predominantly Type II (see Section~\ref{subsec:tempevolgrad}). 

Neon in our PNe shows a comparable behaviour -a statistically significant but moderate negative slope ($-0.027\pm0.004$) and intercept ($8.29\pm0.03$) close to that of oxygen-— consistent with their common nucleosynthetic origin in stars. However, this slope is shallower than those found in some previous studies, such as \citet{Pagomenos2018}, who reported a steeper gradient of $-0.058\pm0.021$ dex kpc$^{-1}$. 
Chlorine and argon also show statistically significant and consistent negative slopes (Cl: $-0.025\pm0.006$, Ar: $-0.027\pm0.004$). Moreover, these gradients remain essentially unchanged when the $t^2$ parameter is included, which reinforces their reliability as tracers of the chemical evolution of the disk. This is consistent with the expectations that Cl and Ar are unaffected by nucleosynthetic processing in PN progenitors or by temperature inhomogeneities within the nebulae, serving as robust metallicity proxies in PNe \citep{DelgadoInglada2014}. These two elements are in excellent agreement with previous PN determinations  \citep{Perinotto2006, Stanghellini2010}. However, the results presented in this work are statistically more robust, owing to our homogeneous set of abundances and our larger sample size, and exhibit smaller and more reliable uncertainties. Also, it should also be emphasized that the distances used in this work are more reliable, as this parameter was highly uncertain in the past.

\subsection{The impact of $t^2$ and dust correction in radial gradients}

While the statistical power of the PN sample (N $\geq$100 for each element) lends weight to the detected slopes, residual biases associated with distance uncertainties and dust depletion can still modulate the intercepts. Temperature inhomogeneities, on the other hand, influence both intercepts and slopes. 

The mean $t^2$ parameter adopted here for our \hii~regions is $0.038\pm0.008$ \citepalias[as determined by][]{MendezAmayo2022}, while the oxygen dust correction reaches up to 0.1 dex. Although these values differ in magnitude, their effects are not directly comparable. The $t^2$ parameter affects each emission line differently, modifying both the intercepts and the slopes of the abundance gradients compared to those obtained from CELs only. According to \citetalias{MendezAmayo2022}, including $t^2$ in O/H abundances derived from CELs increases them by 0.07--0.40 dex, with a mean offset of 0.16 dex. This mean offset is larger than the effect of the dust correction considered in this work. By contrast, the dust correction assumed in this work remains within the typical uncertainties of the O abundances. Therefore, while the global effect of $t^2$ on the oxygen abundance gradient is significant, the impact of dust correction can be considered secondary.

Indeed, it is worth noting that all intercepts and slopes obtained in this work are preserved even when the O-dust correction is not considered (variations remain below $\sim$0.1 dex in the slopes), thus supporting the robustness of our results.

In addition, the O/H panels of Figures~\ref{fig:NewGradCELS} and \ref{fig:NewGradCELS+t2} show that adopting O-RL abundances for \hii~regions leads to a closer agreement with the O gradient traced by Cepheids, as already pointed out by \citetalias{MendezAmayo2022}. 

This highlights that while $t^2$ corrections play a dominant role in shaping the present-day oxygen gradient, dust depletion has only a minor influence within the adopted uncertainties. In the following subsections, we extend this discussion to the gradients derived from PNe, contrasting them with those of \hii~regions to assess the chemical evolution of the Galactic disk across different epochs.

\subsection{Temporal evolution or radial migration in chemical gradients: PNe and H~II regions comparison}
\label{subsec:tempevolgrad}

\begin{figure*}
\centering
\includegraphics[width=0.9\textwidth]{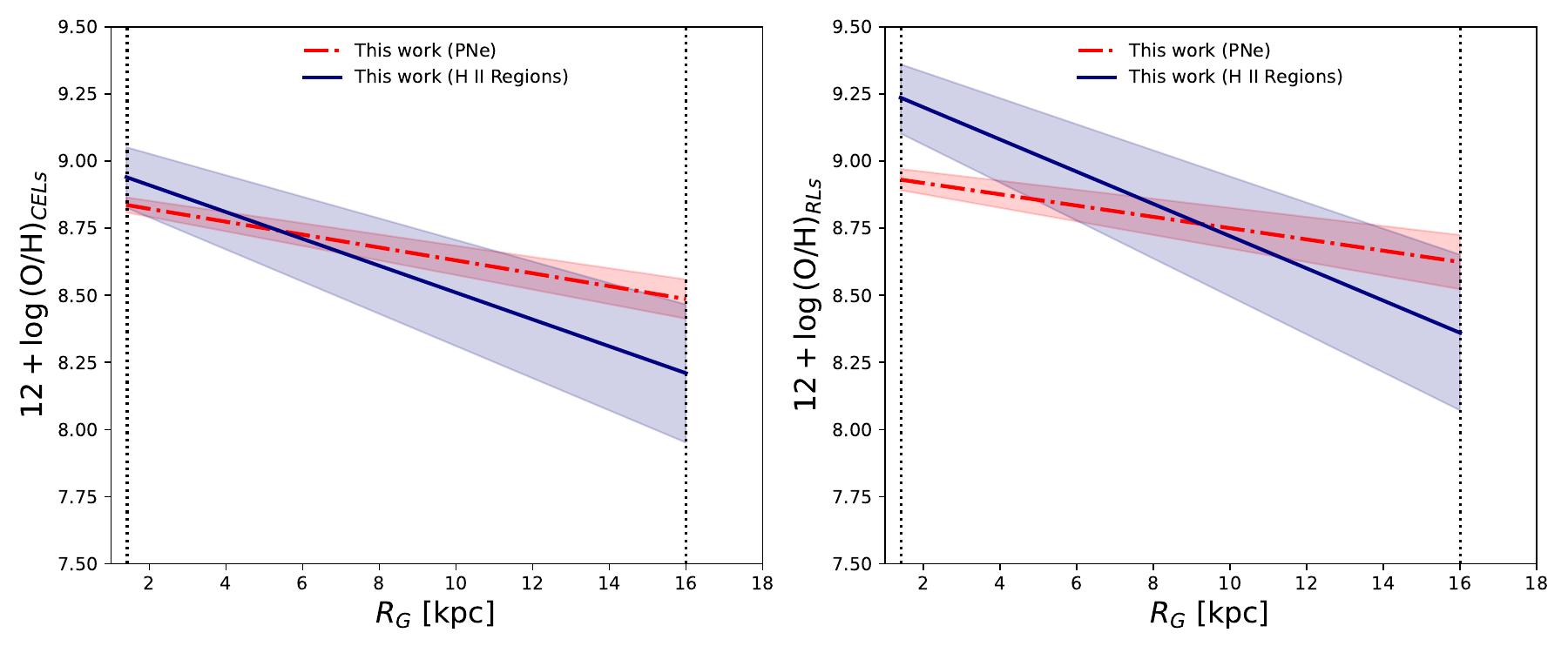}
\caption{Radial abundance gradients of O for PNe (red dashed-pointed line) and for H~II regions (blue solid line) with their respective error bands. Pointed vertical black lines show the 1.42 kpc limit for the bulge and the 16 kpc limit for outer probable halo objects.}
\label{fig:regression_uncertainties}
\end{figure*}

Before comparing the radial gradients derived for PNe and \hii\ regions, it is essential to account for the associated uncertainties. At first glance, Figures~\ref{fig:NewGradCELS} and \ref{fig:NewGradCELS+t2} may suggest that PNe in the outer Galaxy are more metal-rich than \hii\ regions. However, uncertainties increase significantly in these regions, affecting both distance determinations and chemical abundance measurements. To better illustrate this effect, Fig.~\ref{fig:regression_uncertainties} presents the error bands associated with the derived radial gradients. The shaded bands shown in this figure represents the range of linear regression solutions allowed by the uncertainties in both the slope and intercept. As a reminder to the reader, the linear fits were derived through Monte Carlo simulations that account for the uncertainties in both distances and chemical abundances, yielding independent upper and lower uncertainties for each fitting parameter. Using the corresponding upper and lower limits of the slope and intercept, we evaluated the four extreme linear solutions. The envelope defined by the maximum and minimum cases corresponds to the confidence band displayed in the figure, encompassing the full range of physically plausible gradient solutions. Once these uncertainties are properly considered, the apparent differences between the PNe and \hii\ region gradients become less pronounced, indicating that their metallicity distributions may not differ as strongly as initially suggested. In such case, this situation must be accounted for all the following discussion.

Both of our O/H gradients derived from CELs and RLs are flatter in PNe than in \hii\ regions by $-0.03\pm0.01$ dex kpc$^{-1}$ and $-0.04\pm0.01$ dex kpc$^{-1}$, respectively.

The slope of the present day Ne gradient based on CELs is highly uncertain ($\sim$50\%), resulting in a difference of only $-0.01\pm0.02$ dex kpc$^{-1}$ between the gradients derived from \hii~regions and PNe. In the case of the Ne gradient derived from RLs, this difference reaches $-0.03\pm0.02$ dex kpc$^{-1}$. Therefore, to first order, to first order, the Ne gradient results support those obtained for O, although with larger uncertainties. If we average the differences between present-day and past gradients, using both RL- and CEL-based determinations from O and Ne, we obtain a mean value of $-0.028\pm0.008$ dex kpc$^{-1}$ indicating that the gradient traced by older populations is flatter than that traced by younger populations.

Results from Cl and Ar gradients remain inconclusive due to their larger uncertainties. The large uncertainties in these comparisons are mainly driven by the errors associated with the slopes derived for \hii\ regions, which are less precise than those obtained for the PNe.

At face value, this difference could be interpreted as evidence for a temporal steepening of the Galactic abundance gradient. This result is not trivial, as other processes may be affecting the observed trend, and needs a detailed and careful analysis as explained in the following sections. 

\subsubsection{PNe ages}

First, we need to better characterize our PNe sample in order to interpret our results within a temporal framework.

PNe originate from stars spanning a wide range of initial masses (from 0.8$M_{\odot}$ to 8.0$M_{\odot}$). Consequently, their progenitors were formed at different epochs in the past, and it is therefore not strictly correct to treat them as a homogeneous population of the same age.

Some methods have been proposed in the literature to estimate the PN ages, based on different stellar models, proxies, and assumptions used to infer the initial mass ($M_i$) of the progenitor stars and, hence, the age of the PNe. However, stellar age determinations for PNe progenitors are still intrinsically uncertain. Kinematic ages, for example, depend on poorly constrained assumptions about nebular geometry and expansion histories (e.g., constant expansion velocity, spherical morphology, bias towards younger, brighter nebulae in surveys), leading to uncertainties of at least 25\% \citep{Maciel2011, Jacob2013}. Additionally, not all PNe in our sample have reliable or homogeneous kinematic data. Compiling expansion velocities from a single, consistent source would require an extensive and careful re-analysis of heterogeneous literature data, which is beyond the scope of this study.  
Ages derived from post-AGB evolutionary tracks are similarly model-dependent and sensitive to distance, luminosity, and effective temperature, with typical uncertainties of more than $0.2-0.3$ dex and systematic differences between evolutionary models \citep[e.g.,][]{Vassiliadis1994, Frew2008, Bertolami2016}. 

Regardless of this, we carried on an effort to locate our PNe sample in a temporal range, or at least, to know if they are more probably descendants of high- or low- mass progenitors. In such an effort, we explore some of the most popular aging methods in literature, as explained below.  

\begin{figure*}
    \centering
    \subfloat[]{
    \includegraphics[width=0.48\textwidth]{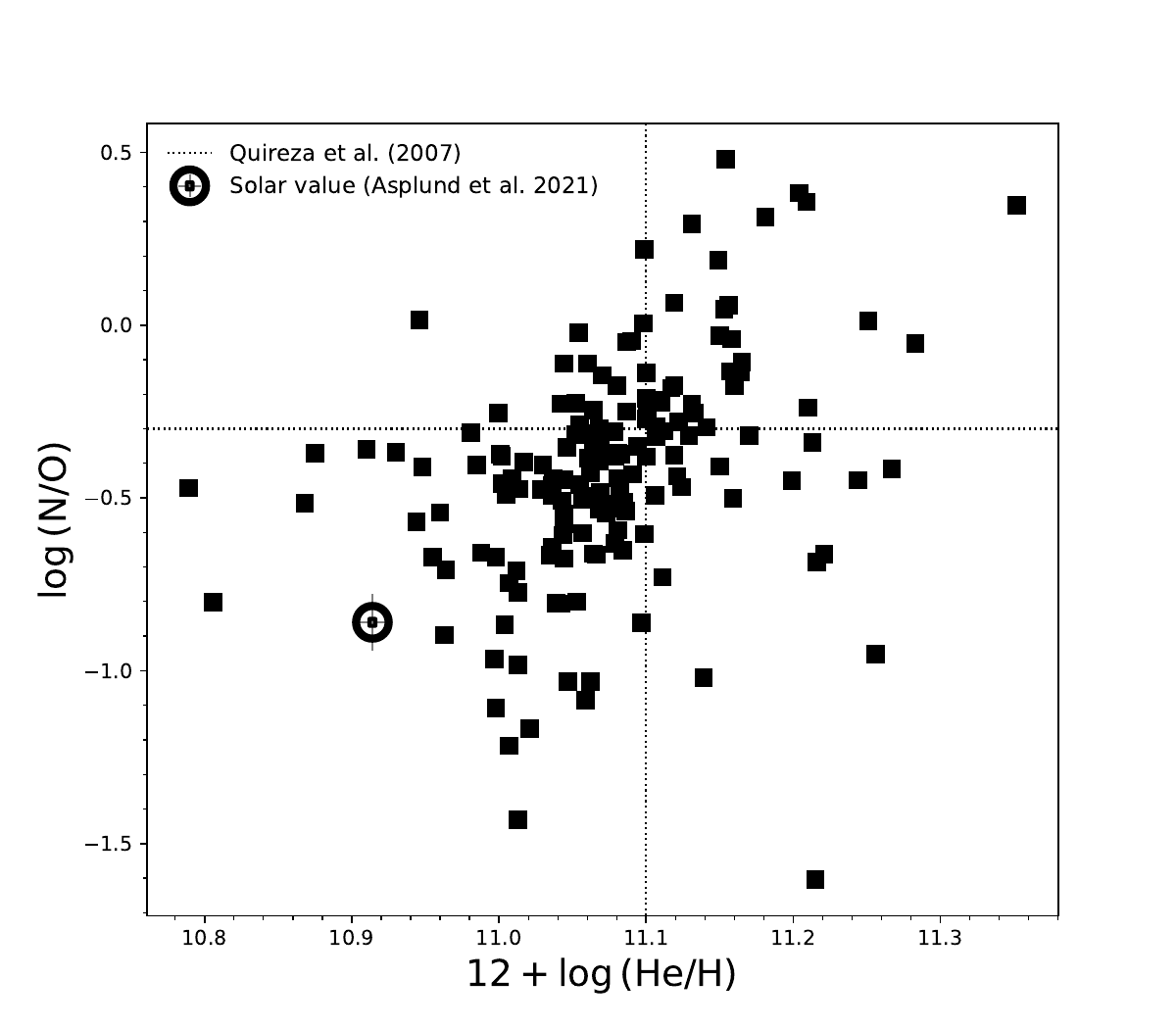}
        \label{fig:HeH_NO}
        }
    \subfloat[]{
        \includegraphics[width=0.48\textwidth]{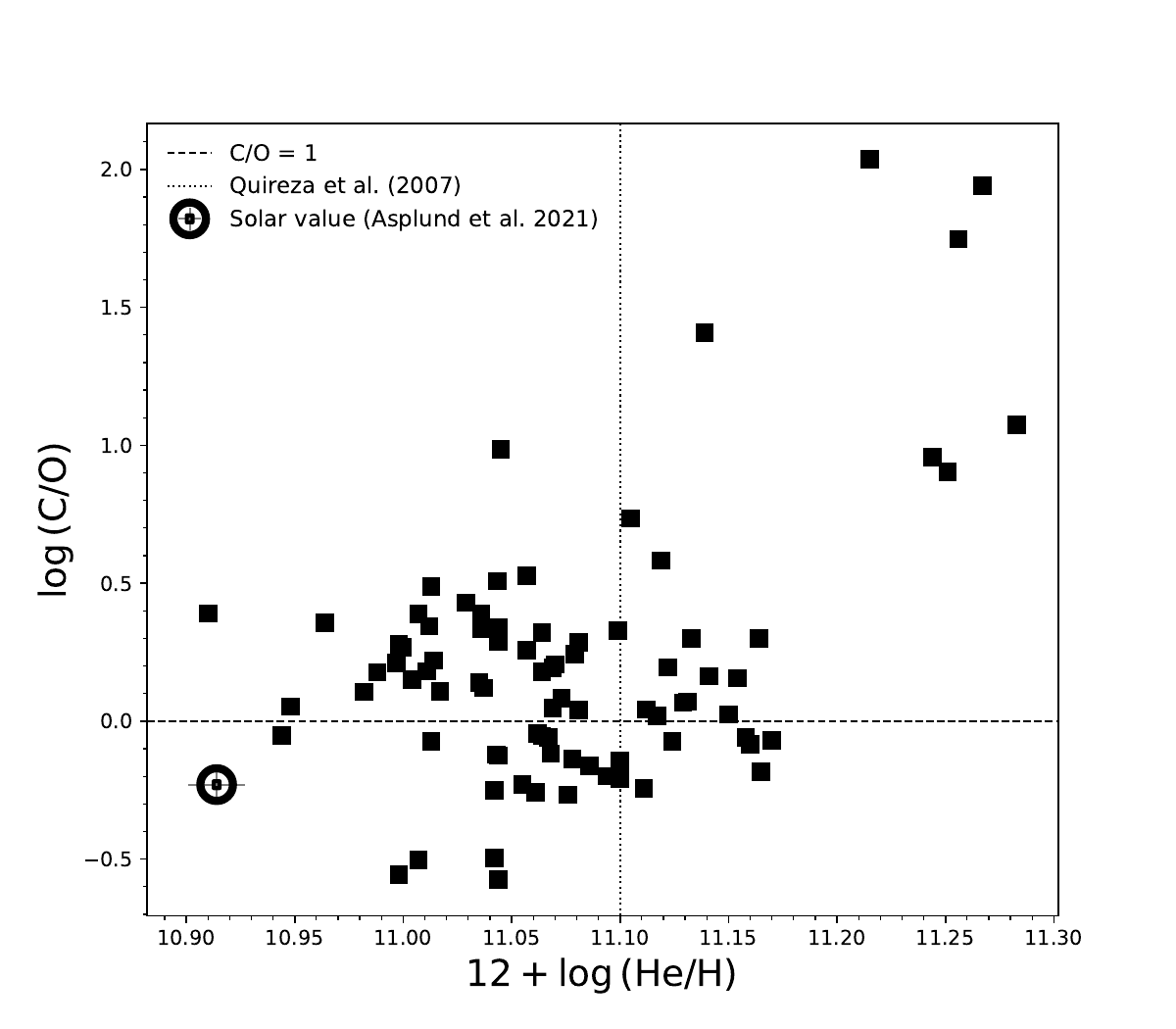}
        \label{fig:HeH_CO}
    }
    \caption{He, C, and N abundances comparison of our PNe sample.}
    \label{fig:NCCl}
\end{figure*}

One method is based on the chemical abundances analysis. When examining the abundance ratios of He/H, N/O, and C/O (see Fig.~\ref{fig:NCCl}) and comparing our PNe distribution with the thresholds proposed by \citet{Quireza2007} to distinguish young PNe (indicated by dotted lines), most of our sample does not appear to be particularly enriched in He or N, and most objects exhibit a C/O ratio greater than 1. These results imply that the majority of our sample (though not all) did not originate from the high-initial mass regime ($M_{i}>4-5 M_{\odot}$). Consequently, these stars did not undergo an efficient hot bottom burning process but instead experienced an efficient third dredge-up. Both features are characteristic of intermediate-initial mass progenitors ($1.5 M_{\odot} < M_{i} < 4-5 M_{\odot}$).

On the other hand, we can apply the \citet{Peimbert1978} classification to categorize our PNe into age groups. According to this classification scheme, Type~I PNe are defined as young, He- and N-rich objects ($2 < M_{i} < 8~M_{\odot}$) with enhanced helium abundances (He/H$>0.125$) and high nitrogen-to-oxygen ratios ($\log$(N/O)$>-0.3$). Type~II objects represent an intermediate-age population ($M_i<2~M_{\odot}$) with no significant He or N enrichment. Finally, Type~III PNe are those located at significant vertical heights ($|z| > 1$~kpc) and exhibiting high peculiar velocities. Using this method, we find that most of our PNe consist of Type~II PNe, regardless of whether CELs or RLs are used (21 Type~I and 118 Type~II PNe based on CEL abundances; 19 Type~I and 120 Type~II based on RLs abundances). This result is consistent with our previous findings regarding initial masses.

A final method to explore the age distribution of our PNe sample is based on the N/O ratio, following the empirical relations proposed by \citet{Cazetta&Maciel2000}, \citet{Maciel2001}, and \citet{Maciel2010}. These expressions allow for the estimation of the progenitor initial masses and their corresponding ages. Using these relations (with both CELs and RLs), our sample is characterized by an average initial mass of $2.04\pm1.4~M_{\odot}$ (with a difference of $0.0079$ dex between CELs and RLs based N/O ratios), which leads to an age of 2.40 Gyr using their corresponding equations. Although this method is subject to significant uncertainties ---ranging from 0.3 and up to 13 Gyr when these are considered--- it provides a first-order estimate of the initial mass range of our sample.

Another method to estimate ages in PNe is based on their kinematic information. However, not all PNe in our sample have reliable or homogeneous kinematic data. Compiling expansion velocities from a single, consistent source would require an extensive and careful re-analysis of heterogeneous literature data, which is beyond the scope of this study \citep[e.g., ][]{Lopez2012}.

The methods explored here suggest that our PNe sample is dominated by intermediate-mass progenitors, with a smaller contribution from more massive progenitors. 
According to \citet{RomanoChiappini2005}, a progenitor star with an initial mass of $2.04\pm1.4~M_{\odot}$ (as obtained with Maciel et al. expressions) corresponds to an age range of $\sim1.60$ Gyr. Thus our PNe sample is probably located (with very high uncertainties considered) in the $1.60-2.4$ Gyr range.

\subsubsection{Comparison with chemical evolution models}

Our gradient derived from our older sample is significantly flatter than those predicted by chemical evolution models, which generally predict either a flattening or a much weaker evolution over time.

For example, \citet{Grisoni2018} only obtains a flatter gradient for older tracers than in younger ones by forcing a variable star formation efficiency \citep[see Fig.6 of][]{Grisoni2018}, while the behaviour is the opposite with a naive constant star formation efficiency.

\begin{figure*}
\centering
\includegraphics[width=0.7\textwidth]{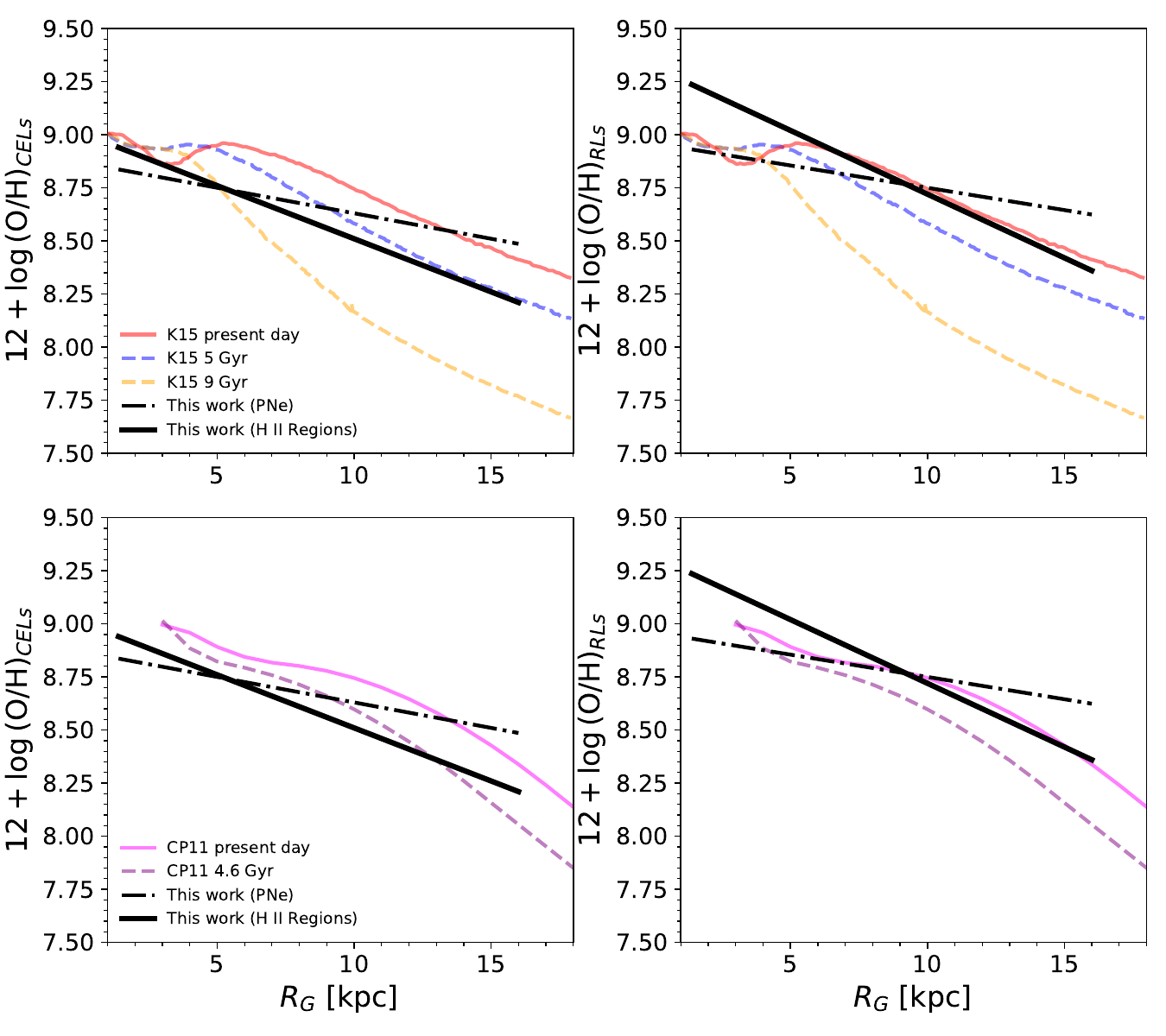}
\caption{Radial gradients of O in the gas phase of ISM, obtained by chemodynamical models of \citet{Kubryk2015} (with radial migration, upper panels) and the chemical evolution models of \citet{Carigi2019} (no radial migration, lower panels), compared to our gradients for \hii\ regions (black solid line) and PNe (dashed-pointed black line) with CELs (left column) and RLs (right column). Models of \citet{Kubryk2015} are shown for three different ages: the present-day (red solid line), for 5 Gyr (blue dashed line) and 9 Gyr ago (yellow dashed line) after the Milky Way formation. \citet{Carigi2019} models are shown for two times: present-day (magenta solid line) and 4.6 Gyr ago (purple dashed line)}. 
\label{fig:K15models}
\end{figure*}

In Fig.~\ref{fig:K15models} we compare our \hii\ regions (black solid line) and PNe (dash-point line) gradients with the chemical evolution models of \citet{Kubryk2015} (for 4, 8, and 12 Gyr after Galactic formation in yellow dashed, blue dashed, and red solid lines, respectively) and \citet{Carigi2019}, (in particular the CP11 model, originally presented in \citealt{Carigi2011}, for 13 Gyr in the magenta solid line and for 8.4 Gyr in the purple dashed line). Left panels show the comparison with our CELs-based gradients, while the right panels show our RLs-based gradients.

As can be seen, all chemical evolution models fail to reproduce our CELs-based gradients. Our current \hii~regions gradient lies below both the \citet{Kubryk2015} and \citet{Carigi2019} models for the present day by $\sim0.75$ dex in both cases. This common result is noteworthy, considering the relevant systematic differences between the two sets of models. On the other hand, our RLs-OH gradient for \hii~regions is reproduced only in a limited radial range in both cases (of 9 to 15 kpc in the case of \citet{Carigi2019} and between 6 to 13 kpc, for \citet{Kubryk2015}), but the comparison still better when temperature fluctuations are taken into account.

Regarding the comparison of our PNe with the models for earlier epochs (dashed lines in all panels), none of the models adequately reproduces our observations. This is because, again, the gradient derived from our older sample is significantly flatter than those predicted by the models.

This result suggests that the PNe-gradient may not trace the intrinsic gas abundance pattern at the time of birth, but rather reflects the dispersion introduced by the radial migration experienced by their progenitor stars during their lifetimes. From the perspective of chemical evolution studies, gradients derived from PNe are not particularly useful for constraining the past abundance gradient of the Galaxy. However, within the framework of chemo-dynamical evolution, they may provide valuable clues to test whether the progenitors of PNe have undergone significant radial migration. Quantifying this migration, however, requires reliable age estimates for the PN progenitors, which remain very uncertain today.

\subsubsection{Comparison with temporal evolution of gradients in literature}

Our findings contrast with those of \citet{Maciel2003, Maciel2006} and \citet{Maciel2007}, who reported a temporal flattening of the oxygen gradient extending up to 9 Gyr in the past (from the past to the present). Although their PN sample is comparable to ours in both size and radial coverage, their chemical abundances were derived from a compilation of heterogeneous literature sources. As the authors themselves noted, at the time of their study, no dataset simultaneously offered both homogeneous abundances and a sample large enough to provide statistically reliable results — a limitation that highlights and distinguishes the present work from what was available in earlier studies.

\citet{Stanghellini2010} in contrast, found indications that the radial metallicity gradients are steeper for younger stellar populations than for older ones, implying a steepening of the gradient over time. Our measured steepening with time is consistent, yet surpasses their work in precision and again, in the homogeneity of the chemical abundances employed.

When compared with extragalactic studies of spiral galaxies, the results show that gradient evolution is not universal. Some galaxies exhibit differences between the gradients of young and old populations (with the former generally being steeper than the latter), whereas others are consistent with little or no change between populations. From the comparison of \hii~regions and PNe gradients, for example, \citet{Magrini2016} (for M~33, M~31, NGC~300 and M~81), \citet{Magrini2010} and \citet{Stasinska2013} (for M~33) found that younger objects tend to show steeper gradients, by $\sim0.1-0.3$ dex kpc$^{-1}$, compared to older PNe. Another work with similar results is that of \citet{Stanghellini2014} (for NGC~300), where the difference found between PNe and \hii~regions gradients is even larger ($\sim-0.05$ to $-0.08$ dex kpc$^{-1}$). In contrast, \citet{Magrini2007} found nearly identical gradients for \hii~regions and PNe gradients, with differences smaller than 0.01 dex kpc$^{-1}$, suggesting no temporal evolution. However, some of these results should be taken with caution, as \citet{DelgadoInglada2020} reported that some PNe reported in M~81 and M~33 by \citet{Magrini2016} can be actually compact \hii~regions, which imply some relevant chemical differences. Furthermore, in all these studies, the dispersion in PNe O/H abundances is significantly higher than in \hii~regions, which suggests the influence of some other processes that introduce noise into the gradients of older objects (e.g., radial migration, modification of O abundances during PNe lifetimes, etc.), such as the ones discussed in this work.

\citet{Arnaboldi2022} studied gradients in M~31 following a different methodology. By employing O and Ar in PNe and \hii~regions and, based on their extinction, Ar abundances and kinematics, they classified PNe into older progenitors (with ages $>4.5$ Gyr) and younger progenitors (with ages $<2.5$ Gyr). In addition to finding a positive gradient for the older population, they consistently find a difference of $\sim0.02$ dex kpc$^{-1}$, with the younger population exhibiting a steeper slope.

Finally, stellar tracers provide important constraints on the temporal evolution of radial abundance gradients. In particular, open clusters and asteroseismic targets have become key tools due to their well-determined ages and distances. 
For example, \citet{Willett2023} analyzed a sample of $\sim600$ clusters spanning $\sim0.1-7$ Gyr and $6–21$ kpc, finding evidence for mild evolution (of $\sim0.03$ dex kpc$^{-1}$ from the youngest stars to stars older than 10 Gyr) over the last few Gyr. Earlier complementary studies also suggest only weak flattening with time or largely preserved gradient shapes. For example, \citet{Magrini2009} obtained little to no significant evolution with time, with slopes remaining roughly constant (e.g., $\sim-0.07$ dex kpc$^{-1}$ for old populations). A similar result is obtained in \citet{Magrini2023}, where open clusters spanning ages from $\sim0.1$ to 7 Gyr were used to trace the Galactic metallicity gradient. They derived an average [Fe/H] slope of $-0.054$ dex kpc$^{-1}$ and found that, after accounting for possible systematic effects, the gradient has undergone only minimal evolution during the last $\sim3$ Gyr, supporting a slow and relatively stationary evolution of the Galactic thin disk. The discrepancy between open clusters' gradient results and ours may arise from differences in tracers and systematics: open clusters can be affected by orbital migration and uncertain birth radii. Besides, our approach (and extragalactic studies) relies on direct comparisons between populations tracing different epochs, which may be more sensitive to subtle evolution. This highlights that the inferred temporal behavior may depend on the specific tracer and methodology employed.

In summary, our findings are consistent with most extragalactic studies where \hii~regions and PNe gradients are compared. However, in all studies (including ours), the effect of migration, mergers, self-contamination, and other factors makes it difficult to be conclusive. It remains challenging to determine whether the results truly reflect a steepening of the gradient over time in the ISM or are driven by these uncertain dynamical and evolutionary factors.

\section{Conclusions}
\label{sec:Conclu}

To investigate the temporal evolution of gradients in the Galactic disk, we compared the O, Ne, Ar, and Cl gradients of 42 Galactic \hii~regions with high-quality spectra \citep[derived by][]{MendezAmayo2022} to those obtained for 176 PNe from the DESIRED catalogue. We computed two sets of abundances for all objects based on the emission lines used to derive them: one from CELs and another obtained by applying a temperature fluctuation parameter ($t^2$) (CELs$+t^2$). 
In addition, we corrected oxygen abundances for dust depletion, carefully reviewing the relevant literature, and compared our results for \hii~regions with the O abundances of young Cepheid stars obtained by \citet{Luck2008}, which are expected to trace the present-day composition of the ISM, similarly to \hii~regions.

Galactocentric distances for our PNe sample were gathered from the careful selection of \citet{HernandezJuarez2024}, who selected the most reliable distance estimate for each PNe among parallaxes-based and statistical determinations. Eleven PNe not included in that work were assigned distances from \citet{Frew2016}, except for M~1-11, whose distance was taken from \citet{BailerJones2021}, based on the Gaia EDR3 association discussed by \citet{ChornayWalton2021} and \citet{GonzalezSantamaria2021}. We excluded PNe likely belonging to the Galactic bulge ($R_{G}=1.45\pm0.02$ kpc), and to the halo (beyond $R_{G}$=17 kpc, and heights above 1 kpc from the Galactic plane), thereby retaining only objects associated with the Galactic disk. After applying these selection criteria, the final PN sample comprises 176 objects. This is the first study in which distances have been incorporated and used for the PNe in the DESIRED sample.

Using the chemical abundances and distances compiled for both \hii~regions and PNe, we compute O, Ne, Cl, and Ar gradients, both from CELs and CELs$+t^2$. We find that all gradients are statistically significant, as indicated by their $p$-values$<$0.05, except for those derived for Cl and Ar in \hii~regions, which should therefore be interpreted with caution.

The O gradient obtained for \hii~regions is consistent with that derived from Cepheid stars, as expected. However, if the presence of internal temperature inhomogeneities in the nebulae is not considered ($t^2=0$), a systematic offset arises that cannot be attributed to oxygen depletion onto dust. Considering $t^2>0$ yields a much better agreement between the distribution of O obtained from \hii~regions and that derived from Cepheid stars.

When comparing the gradients of \hii~regions to those of PNe, we find the following results. 

\begin{itemize}
    \item O gradients obtained with PNe (both with CELs and RLs) have statistically-significant lower slopes than the ones obtained with H~II regions, suggesting a temporal steepening of $-0.03\pm0.01$ dex kpc$^{-1}$. Although modest, this difference exceeds the uncertainties of both slopes.

    \item The Ne gradients are consistent with this result, albeit with larger uncertainties, leading to a steepening of $-0.03\pm0.02$ dex kpc$^{-1}$ with RLs.
    
    \item The average difference of the slopes of \hii~regions and PNe gradients of both O and Ne -CELs and -RLs gradients is statistically significant, of $-0.028\pm0.008$ dex kpc$^{-1}$.

    \item Other $\alpha$-elements show no firm temporal trend when CELs are used. For chlorine and argon, the slopes derived from PNe and \hii~regions are statistically indistinguishable within the errors. 
    
\end{itemize}

Our results are broadly consistent with many extragalactic studies comparing \hii\ regions and PNe, where older populations tend to exhibit flatter gradients, although the magnitude of this effect varies. They also agree with some previous Galactic works that suggest a steepening with time, while differing from others that report flattening or negligible evolution. This diversity highlights that the inferred temporal behavior depends strongly on the tracer, methodology, and underlying systematics.

Studying the abundance trends in our PNe sample and three different aging methods in the literature, we find that most PNe in our sample originate from intermediate-mass progenitors ($\sim2.04\pm1.4$ $M_{\odot}$, corresponding to ages of 1.6–2.4 Gyr), although this estimation is highly uncertain.

The flatter gradients traced by PNe could suggest a temporal steepening of the Galactic abundance gradient in ISM. However, this behavior is not reproduced by chemical evolution models, where the opposite result is obtained \citep[e.g.,][]{Kubryk2015, Carigi2019}. If radial migration is considered, the flatter gradients currently observed in PNe would not directly trace the ISM abundances at the time of their formation, but rather reflect the cumulative effects of the disk’s long-term dynamical evolution.
Rather than limiting their usefulness, this finding emphasizes the potential of PNe as probes of radial migration. Their gradients encode valuable information about the efficiency of stellar redistribution processes, making them important constraints for chemo-dynamical and hydrodynamical models. However, fully exploiting this potential requires improved age determinations, more precise distances, and larger homogeneous samples.

Upcoming surveys such as SDSS-V’s Local Volume Mapper \citep[LVM,][]{Drory2024} will provide hundreds of high-quality IFU spectra of Galactic ionized nebulae, paving the way for systematic and statistically robust studies of abundance gradients across multiple tracers.

\section*{Acknowledgements}

AA thanks SECIHTI for her SNI III-Researcher Assistant grant (CVU No. 825508). 
JEMD and LC thanks the support of the SECIHTI CBF-2025-I-2048 project “Resolving the Internal Physics of Galaxies: From Local Scales to Global Structure with the SDSS-V Local Volume Mapper” (PI: Méndez-Delgado). JG-R acknowledges support from the Agencia Estatal de Investigaci\'on del Ministerio de Ciencia, Innovaci\'on y Universidades (AEI/MCIU) and from the European Regional Development Fund (ERDF) under grants ``The internal structure of ionised nebulae and its effects in the determination of the chemical composition of the interstellar medium and the Universe'' with reference PID2023-151648NB-I00 (DOI:10.13039/501100011033), and ``Planetary nebulae as the key to understanding binary stellar evolution'' with reference PID2022-136653NA-I00 (DOI:10.13039/501100011033).
The authors thank Dr. Gloria Delgado Inglada for her valuable support, suggestions and scientific discussions in developing this work. 

\section*{Data Availability}

All data used in this work may be found in the tables of the paper, both in the main body and in the Appendix section.





\bibliographystyle{mnras}
\bibliography{Mwchemevol}

\section{Appendix}
This section contains all data used in the present work. Table~\ref{tab:PNerefs} shows the original references of the spectral data for each PNe. Table~\ref{tab:PNeCELsabunds1} presents the DESIRED PNe sample CELs-based ($t^2=0$) abundances, while Table~\ref{tab:PNeRLsabunds1} contains the RLs-based ($t^2>0$) abundances. Table~\ref{tab:PNedists} shows the heliocentric distances obtained from the heliocentric distances of \citet{HernandezJuarez2024}, along with their coordinates in degrees. Finally, Table~\ref{tab:PNeages} shows ages obtained with the methods explored in this work, obtained from different chemical abundances ratios. All tables are available to be sent in electronic format via private communication with the authors.  

\onecolumn

\begin{longtable}{|l|c|l|}
\caption{References for DESIRED PNe used in this work. These correspond only to objects with "PN" and "Milky Way" tags in the complete DESIRED catalogue.}\label{tab:PNerefs}
 \\
\hline
\hline
NAME & PNG & Reference \\
\hline
\endfirsthead
\caption{Continued.}\\
\hline
NAME & PNG & Reference \\
\hline
\endhead
\hline
\endfoot
\hline
\endlastfoot
M2-23 & 002.2-02.7 & \citet{WangLiu2007}\\
Cn1-5 & 002.2-09.4 & \citet{GarciaRojas2012}\\
NGC6369 & 002.4+05.8 & \citet{GarciaRojas2012} \\
M1-38 & 002.4-03.7 & \citet{Milingo2002} \\
M1-42 & 002.7-04.8 & \citet{GarciaRojas2022} \\
Hb4 & 003.1+02.9 & \citet{GarciaRojas2012} \\
M2-36 & 003.2-06.2 & \citet{Espiritu21} \\
NGC6565 & 003.5-04.6 & \citet{Milingo2002} \\
\hline
\end{longtable}

\begin{longtable}{|l|c|c|c|c|c|c|c|}
\caption{CELs-based abundances ($t^2=0$) gathered in this work for DESIRED PNe [in the usual format of $12+\log$(X/H)]. O is corrected by 0.1 dex for dust.} \\
\hline
\hline
PNe & PNG & ${\rm He}_{\rm CELs}$ & ${\rm O}_{\rm CELs}$ & ${\rm N}_{\rm CELs}$ & ${\rm Ne}_{\rm CELs}$ & ${\rm Cl}_{\rm CELs}$ & ${\rm Ar}_{\rm CELs}$\\
\hline
\endfirsthead
\caption{Continued.}\\
\hline
PNe & PNG & ${\rm He}_{\rm CELs}$ & ${\rm O}_{\rm CELs}$ & ${\rm N}_{\rm CELs}$ & ${\rm Ne}_{\rm CELs}$ & ${\rm Cl}_{\rm CELs}$ & ${\rm Ar}_{\rm CELs}$\\
\hline
\endhead
\hline
\endfoot
\hline
\endlastfoot
\label{tab:PNeCELsabunds1}
M2-23 & 002.2-02.7 & $11.042_{-0.016}^{+0.019}$ & $8.445_{-0.029}^{+0.030}$ & $8.217_{-0.447}^{+0.159}$ & $7.730_{-0.043}^{+0.041}$ & $4.417_{-0.488}^{+0.160}$ & $5.575_{-0.094}^{+0.099}$ \\
Cn1-5 & 002.2-09.4 & $11.164_{-0.018}^{+0.017}$ & $8.883_{-0.031}^{+0.030}$ & $8.746_{-0.104}^{+0.067}$ & $8.696_{-0.043}^{+0.047}$ & $5.461_{-0.062}^{+0.057}$ & $6.658_{-0.054}^{+0.055}$ \\
NGC6369 & 002.4+05.8 & $11.036_{-0.016}^{+0.015}$ & $8.638_{-0.031}^{+0.029}$ & $8.174_{-0.049}^{+0.047}$ & $8.020\pm0.039$ & $5.230_{-0.062}^{+0.066}$ & $6.372_{-0.055}^{+0.056}$ \\
M1-38 & 002.4-03.7 & -- & $8.871_{-0.275}^{+0.280}$ & $8.091_{-0.571}^{+0.184}$ & -- & -- & $5.998_{-0.175}^{+0.170}$ \\
M1-42 & 002.7-04.8 & $11.221_{-0.019}^{+0.020}$ & $9.112_{-0.052}^{+0.047}$ & $8.449_{-0.100}^{+0.068}$ & -- & $5.348_{-0.054}^{+0.055}$ & $6.574_{-0.040}^{+0.041}$ \\
Hb4 & 003.1+02.9 & $11.070_{-0.025}^{+0.023}$ & $8.783_{-0.038}^{+0.039}$ & $8.638_{-0.115}^{+0.074}$ & $8.204_{-0.058}^{+0.055}$ & $5.402_{-0.136}^{+0.149}$ & $6.599_{-0.124}^{+0.112}$ \\
M2-36 & 003.2-06.2 & $11.119_{-0.006}^{+0.007}$ & $8.840\pm0.019$ & $8.904_{-0.043}^{+0.038}$ & $8.505_{-0.022}^{+0.023}$ & $5.674_{-0.032}^{+0.026}$ & $6.804_{-0.016}^{+0.015}$ \\
NGC6565 & 003.5-04.6 & $11.061_{-0.025}^{+0.026}$ & $8.834_{-0.047}^{+0.044}$ & $8.449_{-0.139}^{+0.082}$ & $8.212\pm0.074$ & $5.370_{-0.068}^{+0.072}$ & $6.491_{-0.062}^{+0.065}$ \\
\hline
\end{longtable}

\begin{longtable}{|l|c|c|c|c|c|c|c|}
\caption{RLs-based abundances ($t^2>0$) gathered in this work for DESIRED PNe [in the usual format of $12+\log$(X/H)]. O is corrected by 0.1 dex for dust.} \\
\hline
\hline
PNe & PNG & ${\rm He}_{\rm RLs}$ & ${\rm O}_{\rm RLs}$ & ${\rm N}_{\rm RLs}$ & ${\rm Ne}_{\rm RLs}$ & ${\rm Cl}_{\rm RLs}$ & ${\rm Ar}_{\rm RLs}$\\
\hline
\endfirsthead
\caption{Continued.}\\
\hline
PNe & PNG & ${\rm He}_{\rm RLs}$ & ${\rm O}_{\rm RLs}$ & ${\rm N}_{\rm RLs}$ & ${\rm Ne}_{\rm RLs}$ & ${\rm Cl}_{\rm RLs}$ & ${\rm Ar}_{\rm RLs}$\\
\hline
\endhead
\hline
\endfoot
\hline
\endlastfoot
\label{tab:PNeRLsabunds1}
M2-23 & 002.2-02.7 & $10.881_{-0.078}^{+0.057}$ & $7.808\pm0.149$ & $7.531_{-0.503}^{+0.155}$ & $6.989_{-0.189}^{+0.171}$ & $4.254_{-0.457}^{+0.155}$ & $5.554_{-0.084}^{+0.092}$ \\
Cn1-5 & 002.2-09.4 & $11.157_{-0.019}^{+0.018}$ & $9.254_{-0.047}^{+0.044}$ & $9.107_{-0.096}^{+0.060}$ & $9.095_{-0.064}^{+0.066}$ & $5.528_{-0.059}^{+0.060}$ & $6.720_{-0.057}^{+0.056}$ \\
NGC6369 & 002.4+05.8 & $11.023_{-0.019}^{+0.018}$ & $8.429_{-0.046}^{+0.048}$ & $7.969_{-0.079}^{+0.060}$ & $7.771_{-0.066}^{+0.059}$ & $5.178_{-0.060}^{+0.059}$ & $6.353_{-0.049}^{+0.047}$ \\
M1-38 & 002.4-03.7 & -- & $8.865_{-0.278}^{+0.257}$ & $8.112_{-0.577}^{+0.177}$ & -- & -- & $5.862_{-0.174}^{+0.170}$ \\
M1-42 & 002.7-04.8 & $11.206_{-0.020}^{+0.021}$ & $9.351_{-0.049}^{+0.051}$ & $8.690_{-0.105}^{+0.076}$ & -- & $5.376_{-0.053}^{+0.056}$ & $6.634_{-0.044}^{+0.039}$ \\
Hb4 & 003.1+02.9 & $11.068_{-0.023}^{+0.022}$ & $9.089_{-0.097}^{+0.093}$ & $8.950_{-0.122}^{+0.076}$ & $8.597_{-0.119}^{+0.113}$ & $5.459_{-0.150}^{+0.146}$ & $6.617_{-0.132}^{+0.127}$ \\
M2-36 & 003.2-06.2 & $11.100_{-0.012}^{+0.010}$ & $8.190_{-0.031}^{+0.030}$ & $8.259_{-0.044}^{+0.038}$ & $7.767_{-0.037}^{+0.042}$ & $5.525\pm0.027$ & $6.751_{-0.015}^{+0.014}$ \\
NGC6565 & 003.5-04.6 & $11.056\pm0.024$ & $9.017_{-0.056}^{+0.053}$ & $8.641_{-0.118}^{+0.086}$ & $8.504_{-0.083}^{+0.094}$ & $5.390_{-0.075}^{+0.079}$ & $6.522_{-0.063}^{+0.066}$ \\
\hline
\end{longtable}

\begin{longtable}{|l|c|c|c|c|c|c|c|}
\caption{Galactocentric distances for DESIRED PNe, compiled from \citet{HernandezJuarez2024}.} \\
\hline
\hline
NAME & PNG & RA [deg] & DEC [deg] & l [deg] & b [deg] & $R_G$ [kpc] & $Z$ [kpc] \\
\hline
\endfirsthead
\caption{Continued.}\\
\hline
NAME & PNG & RA [deg] & DEC [deg] & l [deg] & b [deg] & $R_G$ [kpc] & $Z$ [kpc] \\
\hline
\endhead
\hline
\endfoot
\hline
\endlastfoot
\label{tab:PNedists}
M2-23 & 002.2-02.7 & 270.427672 & -28.428921 & 2.226554 & -2.785081 & $3.225\pm1.977$ & $-0.243\pm0.097$ \\
Cn1-5 & 002.2-09.4 & 277.298542 & -31.499760 & 2.291339 & -9.480634 & $3.748\pm1.675$ & $-0.758\pm0.296$ \\
NGC6369 & 002.4+05.8 & 262.335236 & -23.759659 & 2.432014 & 5.846860 & $7.068\pm0.467$ & $0.116\pm0.047$ \\
M1-38 & 002.4-03.7 & 271.523920 & -28.675004 & 2.483100 & -3.745380 & $1.887\pm2.447$ & $-0.418\pm0.170$ \\
M1-42 & 002.7-04.8 & 272.770790 & -28.983080 & 2.738620 & -4.851350 & $3.845\pm1.666$ & $-0.372\pm0.144$ \\
Hb4 & 003.1+02.9 & 265.470000 & -24.702300 & 3.170790 & 2.928473 & $5.313\pm1.196$ & $0.148\pm0.061$ \\
M2-36 & 003.2-06.2 & 274.422625 & -29.138860 & 3.282810 & -6.198800 & $2.182\pm2.223$ & $-0.669\pm0.270$ \\
NGC6565 & 003.5-04.6 & 272.968691 & -28.178450 & 3.531703 & -4.621267 & $4.338\pm1.574$ & $-0.314\pm0.129$ \\
\end{longtable}

\begin{longtable}{|l|c|c|c|c|c|}
\caption{Ages estimated with different methods for DESIRED PNe. The first and second columns show the name and PNG identifier of the PNe. The third column shows the C/O ratio, fourth column contains the Peimbert Types obtained by using \citet{Quireza2007} limits on He/H and N/O abundances; fourth and fifth columns show initial progenitor masses and ages computed based on their $\log$(N/O) ratio, using \citet{Cazetta&Maciel2000, Maciel2001, Maciel2003} and \citet{Maciel2010} equations.} \\
\hline
\hline
PNe & PNG & $\log$(C/O) & Peimbert Type & $M^{*}_{i}$(M10) [$M_{\odot}$] & Age [Gyr] \\
\hline
\endfirsthead
\caption{Continued.}\\
\hline
PNe & PNG & $\log$(C/O) & Peimbert Type & $M^{*}_{i}$(M10) [$M_{\odot}$] & Age [Gyr] \\
\hline
\endhead
\hline
\endfoot
\hline
\endlastfoot
\label{tab:PNeages}
M2-23 & 002.2-02.7 & -0.497000 & Type IIa & 2.310777 & 0.664953 \\
Cn1-5 & 002.2-09.4 & 0.299000 & Type I & 2.724818 & 0.419086 \\
NGC6369 & 002.4+05.8 & 0.335000 & Type IIa & 1.834196 & 1.321293 \\
M1-38 & 002.4-03.7 & -- & -- & 1.224457 & 4.910628 \\
M1-42 & 002.7-04.8 & -- & Type IIa & 1.450215 & 2.785505 \\
Hb4 & 003.1+02.9 & 0.205000 & Type IIa & 2.677835 & 0.439518 \\
M2-36 & 003.2-06.2 & 0.582000 & Type I & 4.574858 & 0.114483 \\
NGC6565 & 003.5-04.6 & -0.259000 & Type IIa & 1.986630 & 1.036698 \\
\hline
\end{longtable}

\bsp	
\label{lastpage}
\end{document}